\documentclass[conference,compsoc]{IEEEtran}

\usepackage{hyperref}
\ifCLASSOPTIONcompsoc
 \usepackage[nocompress]{cite}
\else
 \usepackage{cite}
\fi
\ifCLASSINFOpdf
 \usepackage[pdftex]{graphicx}
\else
 \usepackage[dvips]{graphicx}
\fi
\usepackage{amsmath}
\usepackage{algorithmic}

\usepackage{array}

\ifCLASSOPTIONcompsoc
 \usepackage[caption=false,font=footnotesize,labelfont=sf,textfont=sf]{subfig}
\else
 \usepackage[caption=false,font=footnotesize]{subfig}
\fi

\usepackage{stfloats}
\usepackage{url}

\begin{document}
%
\title{Graph Neural Networks for Model Recommendation using Time Series Data}

\author{\IEEEauthorblockN{
Aleksandr Pletnev\IEEEauthorrefmark{4}\IEEEauthorrefmark{1},
Rodrigo Rivera-Castro\IEEEauthorrefmark{4}\IEEEauthorrefmark{2} and
Evgeny Burnaev\IEEEauthorrefmark{3}
}
\IEEEauthorblockA{Skoltech, Moscow, Russia\\
\IEEEauthorrefmark{4}Equal Contribution
\\
Email: \IEEEauthorrefmark{1}aleksandr.pletnev@skoltech.ru,
\IEEEauthorrefmark{2}rodrigo.riveracastro@skoltech.ru,
\IEEEauthorrefmark{3}e.burnaev@skoltech.ru
}
}

\maketitle

\begin{abstract}
Time series prediction aims to predict future values to help stakeholders make proper strategic decisions. This problem is relevant in all industries and areas, ranging from financial data to demand to forecast. However, it remains challenging for practitioners to select the appropriate model to use for forecasting tasks. With this in mind, we present a model architecture based on Graph Neural Networks to provide model recommendations for time series forecasting. We validate our approach on three relevant datasets and compare it against more than sixteen techniques.
Our study shows that the proposed method performs better than target baselines and state of the art, including meta-learning.
The results show the relevancy and suitability of GNN as methods for model recommendations in time series forecasting.

\end{abstract}


%
\IEEEpeerreviewmaketitle

\section {Background} \label{I_background}
Practitioners and researchers require time series prediction in many situations, such as deciding to build another factory over the next five years, planning for call center staff, or stockpile forecasting. The forecasts may take several years in advance (in case of capital investment) or just a few minutes (in case of telecommunication routing). Whatever the circumstances or time frame, forecasting is an essential aid to effective and efficient planning.
Modern time series analysis forces the researcher to consider a wide range of data characteristics. Examples of this are whether the variables are stationary or not, whether each has a trend, and how many lags we should consider when examining data. As a result, this time series data analysis forces the researcher to consider a wide variety of potential models.
The wide variety of models, use cases, and types of time series data available generate a need for an adequate model selection approach. The principal idea of model selection is to estimate the performance of different model candidates to choose the best model achievable, \cite{definition_model}. Among its various benefits are more reliable models and better predictions.

\section{Motivation} \label{I_motivation}
There are two alternative strategies for making so many forecasts: 1) either use one forecasting method for all-time series; or 2) select a suitable forecasting method for each time series separately. It is highly unlikely that there will be the best performing model suitable for the all-time series. Therefore there needs to be an approach for selecting a model for individual time-series. 
There are several general approaches to model selection. The first existing approach is to select a class of models in advance and then select a model in the corresponding class based on the statistical properties. It is usually impossible to compare models from different classes based on statistical properties. Expert judgment is crucial in selecting the most appropriate class of models to use. However, this approach is not generalizable.
The second existing approach relies on generating as many features as possible, given new time-series, and then using a classifier to select the best model based on corresponding features. The problem is that features do not always accurately describe the underlying time-series. Therefore the classifier is not able to consistently select the best model for the associated time-series. 
The third existing approach relies on applying time-series cross-validation, where models from many different classes may be applied. Afterward, the model with the lowest cross-validated MSE is selected. Nevertheless, this approach increases the computation time involved substantially. Also, the approach does not capture underlying information between different time-series, which should be relevant for providing prediction and recommendation.

The subject of this research is to improve the forecasting of the future values of the entity based on previously observed values. 
This work aims to develop the model recommendation approach: a model that captures the underlying structure and relationship between time-series, thus providing relevant time-series model prediction recommendations. 

\section{Innovation} \label{I_innovation}
This work uses the canonical Graph Neural Network architecture for time-series data and serves as an engineering novelty. The theoretical novelty resides in a new algorithm capable of taking time-series data for training and given a new time-series to provide a graph of similarities between different time series. We use this output to provide a model recommendation. The motivation is that time series sharing similarities can use the same forecasting model. Thus, this work provides a novel application as practitioners can use it and save time evaluating models for forecasting and still get good results.

\section{Literature Review}
Model selection methods attempt to choose one of these function classes to avoid both overfitting and underfitting.
\cite{guyon_model_selection} defines model selection as an ensemble of techniques used to select a model that best explains or predicts some future data.

There have been many overview papers on model selection scattered in the communities of signal processing, \cite{signal_processing}, statistics, \cite{statistics}, machine learning, \cite{MachineLearning}, epidemiology, \cite{Epidemiology}, chemometrics, \cite{Chenomotrics}, ecology and evolution, \cite{EcoEvo}.

\paragraph{Akaike information criterion}\label{sec:aic}
AIC is a model selection method proposed by \cite{akaiki_15}. The goal is to approximate the out-sample prediction loss by the sum of the in-sample prediction loss and a correction term. In the typical case where the loss is logarithmic, the AIC approach should select the model, which minimizes
\begin{equation*}
AIC_q = -2 *\hat{t}_{n,q} + 2d_q
\end{equation*}, 

where $\hat{t}_{n,q} $ is the maximized log-likelihood of model given $n$ observations and $d_q$is the dimension of model. 

\paragraph{Bayesian information criterion (BIC)}
Another popular selection method based on information criteria is \cite{bic_21} and is defined as
\begin{equation*}
BIC_q = -2 \hat{t}_{n,q}+d_q *log(n)
\end{equation*}
The difference between AIC is in replacing the logarithm with the sample size.

For our study, we will perform the baseline analysis using AIC, since the datasets we are using do not have the small sample size, and our experiments showed that there is usually little to no difference between models selected using AIC or BIC. 

\paragraph {Cross-validation} 
Machine learners use cross-validation (CV) widely, \cite{CV_38},\cite{CV_39}. CV does not assume that the candidate models are parametric and can be judged by any metric such as MSE.

However, it is necessary to consider the dependence of time series data on time: if we collect a dataset during one session and randomly split it into k parts or a random sample is held for cross-validation, then close or adjacent data will be part of training data. Since neighboring observations are not independent, this leads to an overestimated model effectiveness. This phenomenon is well known in statistics (\cite{Hart}, \cite{Burman}, \cite{Arl}). 
In this research, we use a block study design adapted to the scheme of block cross-validation \cite{Grimes}, which uses whole adjacent sequences of observations for model learning, selection, and validation.

\paragraph {Meta-learning}
Meta-learning is an approach that can be intuitively described as applying machine learning to the algorithm and hyperparameter selection (\cite{ML1},\cite{ML2},\cite{ML3}).
The input data are different machine learning tasks and datasets. The output is a well-performing algorithm and a hyperparameter combination. In meta-learning, we learn "meta-features" to identify similar problems for which an algorithm and hyperparameter combination is right for. These meta-features can include the number of data points, features, classes, the data skewness, the entropy of the targets, and more. 

Although the approach is straightforward, still, there is no extensive research on which approach feature generation approach, classifier to use. For the baseline analysis in our study, we will use the libraries tsfresh and RandomForest. 

\paragraph {Graph-based approaches}
It is possible to create a graph from a non-graph classification dataset using the methods proposed in \cite{GNN_survey} and redefine the model selection task as graph classification and then solve classification tasks. One approach, \cite{GNN_1}, is to try different values for knn to connect the points into a graph, then to obtain a graph for each parameter setting, and to verify the performance of classification of a graph model on the obtained graph. 

\paragraph{Penalty which depends on the model class}
It measures the capacity of a class of models to overfit and penalizes all models in that class accordingly, regardless of their individual properties, \cite{vapnik}.

\paragraph{The method of sieves}
It directly optimizes the fit, but within a constrained class of models and relaxes the constraint as the amount of data grows. 

\paragraph{Encompassing models}
The sampling distribution of any estimator of any model class is a function of the true distribution. If the model class has been well-estimated, it should be able to predict what other, wrong model classes will estimate, but not vice versa. 

\paragraph{Model averaging}
It does not try to pick the best or correct model. Model averaging uses them all with different weights and serves as a smoother.

\paragraph{Adequacy testing}
The correct model should be able to encode the data as uniform IID noise. This technique tests whether "residuals," in the appropriate sense, are IID uniform. It may be possible that none of the models on offer is adequate or models make specific probabilistic assumptions.

\section {Datasets} \label{datasets_description}
We tested the methods on two types of datasets, a commercial and public dataset from the finance market with a daily period. For each dataset, we designed a relationship graph, one of the inputs for our model. 

\begin{table}
\centering
\caption{\textsc{Overview of evaluated Datasets}\label{Tab: datasets}}
\resizebox{0.48\textwidth}{!}{%
\begin{tabular}{lcccc}
\hline
Dataset  & \# Users & \# Relation Types & \# Relations & Time Period \\ \hline
Commercial & 1000    & 200         & 5237      & 365     \\ \hline
NASDAQ   & 972    & 112         & 5745      & 1245     \\ \hline
NYSE    & 1617    & 130         & 21158      & 1245
\\ \hline
\end{tabular}
}
\end{table}

Choco Communications GmbH provides the first dataset. Each time series represents daily ordered gross merchandise volume (GMV) for a particular restaurant. We selected one thousand restaurants and two hundred suppliers with the highest order frequency, respectively. A mutual supplier creates a relationship between two restaurants. On average, restaurants have five suppliers.

The second dataset is devoted to financial markets. We select two markets from 2013 to 2017 for their representative properties that NASDAQ is more volatile, whereas NYSE is more stable, \cite{finmark}.
We perform filtering and preprocessing by retaining stocks satisfying two conditions: (1) The company was traded at the end of the period. 2) The company did not have more than 5\% of missing values during the analyzed period. 
Similarly, we collected the hierarchy structure of NASDAQ and NYSE stocks from the official company list maintained by NASDAQ and extract industry relations for each stock pair under the same industry node.

\section {Model evaluation} \label{evaluation_metrics}
\subsection{Model recommendation labelling} \label{recommendation_labeling}

\begin{figure}[ht]
  \centering
  \includegraphics[width=\columnwidth]{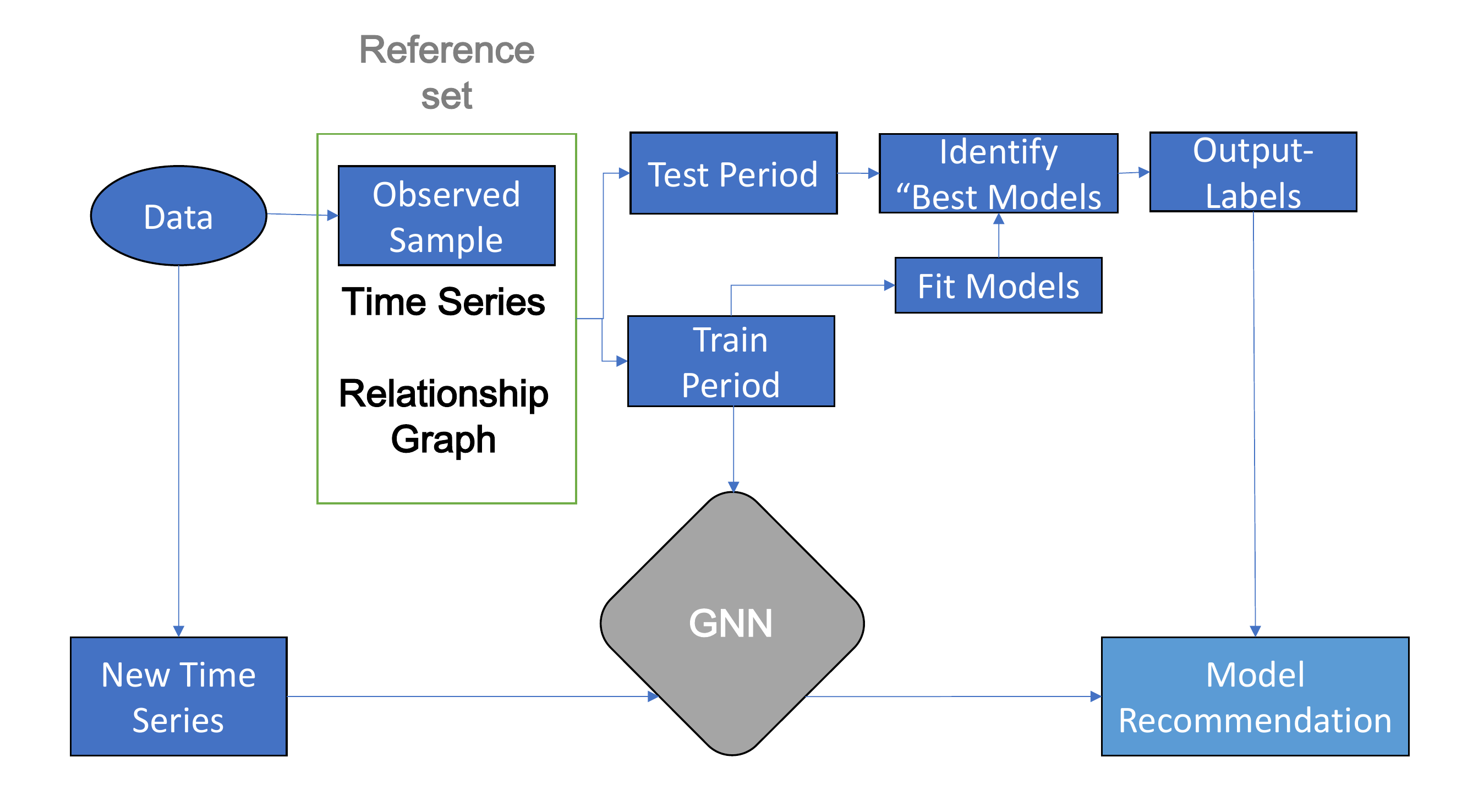}
  \caption{Model Recommendation Labelling}
  \label{fig:recommendation_labelling}
\end{figure}

First, we identify the "best" model for each of the time series. To create labels with the best models, we need an extensive collection of time series, similar to those we will be forecasting. This framework is presented in \autoref{fig:recommendation_labelling}. We can assume that we have an infinite population of time series, and we sample from them to train the classification algorithm denoted as the "observed sample." We wish to forecast new time series as additional draws from the same population. For this, we split each time series into a training period and a test period. Since we are using Time Series Nested Cross-Validation, each time series will have several training periods and testing periods. Then we average the error metric for every testing period for the given time series to identify the "best" model. We label each time series with the model, which has given the lowest error score.

\subsection{Forecasting methods} \label{forecasting_methods}

\paragraph{Naive}
We set all forecasts to be the value of the last observation, \cite{Forecasting3}. That is, $\hat{y}_{T_h|T}=y_T$. This model is optimal when data follows a random walk.
\paragraph{Simple Average} 
Here, the forecasts of all future values are equal to the average historical data, \cite{Forecasting3}, defined as
\begin{equation*}
    \hat{y}_{T+h|T} = \frac{(y_1+...+y_T)}{T}
\end{equation*}

\paragraph{ARIMA family models}
One of the most popular and frequently used stochastic time series models is the Autoregressive  Integrated  Moving  Average  (ARIMA), \cite{box_jen}. The underlying assumption in this model is that the considered time series is linear and follows a particular known statistical distribution, such as the normal distribution. 
We will be using multiple variations of ARIMA (p, d, q) such as white noise ARIMA(0,0,0), Autoregression ARIMA (p,0,0), Moving Average ARIMA(0,0,q), Random walk with drift ARIMA (0,1,0) with a constant and no constant, and canonical ARIMA.

\paragraph{Exponential Smoothing models}
Forecasts produced using exponential smoothing methods (ETS), \cite{ES1}, \cite{ES2}, \cite{ES3}, are weighted averages of past observations, with the weights decaying exponentially as observations age. We consider multiple variations of ETS, such as ETS without trend and seasonal components, ETS with an additive trend and without seasonal component, ETS with an additive damped trend and without seasonal component, ETS with an additive trend and seasonal component (Holt-Winters').
  
\paragraph{Regression approaches}
Statistical methods systematically outperform machine learning methods for univariate time series forecasting \cite{ML_los}. However, machine-learning algorithms make it possible to find patterns in the time series. We consider classic machine learning methods such as, Random Forest Regression, \cite{RForest}, XGBoost Regression, \cite{Xbog}, and Gradient Boosting Regression, \cite{GBR1},\cite{GBR2}.

\paragraph{Neural Network model} 
The Recurrent Neural Network (RNN), \cite{LSTM0},  has achieved solid performance to process sequential data, \cite{LSTM1}, \cite{LSTM2}, \cite{LSTM3}. Among various RNN models, such as vanilla RNN, LSTM, and GRU, \cite{GRU}, we choose LSTM owing to its ability to capture both short-term and long-term dependency.

\subsection{Metrics} \label{metrics}

\paragraph{Symmetric Mean Absolute Percentage Error}
We selected SMAPE as it represents the percentage of average absolute error occurred and gives less significance to outliers since it is a relative metric. Similarly, SMAPE is independent of the scale, but affected by data transformation, \cite{SMAPE1}, \cite{SMAPE2}, \cite{SMAPE3}. We can calculate it as follows: 

$$SMAPE = \frac{\sum_{t=1}^{n}(F_t-A_t}{\sum_{t=1}^{n} (A_t+F_t) },$$ where $A_t$ is the actual value and $F_t$ is the forecast value. 

\paragraph{Mean Squared Error}
MSE penalizes extreme errors that occurred while forecasting. Significant errors are more expensive than small errors. MSE is a good measure of overall forecast error. It is not as intuitive and easily interpretable as the relative metric. We calculate as follows:

$$ MSE = \frac{1}{N}*\sum_{i=1}{N}(A_t-F_t)^2, $$
where $A_t$ is the actual value and $F_t$ is the forecast value. 

\subsection {Baselines} \label{baselines}
We compare the algorithm with two types of baselines. The first is the accuracy of the proposed model compared against other model selection techniques. For the second, we assess the accuracy of the proposed model in the next value forecasting compared to the forecasting approaches.
For the first baseline, we compare our model with the following approaches: Random Model Recommendation, Meta-Learning, \cite{ML1}, \cite{ML2}, \cite{ML3}, Cross-validation, \cite{CV_38}, \cite{CV_39}, and Probabilistic model selection, AIC, see \autoref{sec:aic}.
For the second type of the baseline, we compare the forecasting accuracy of our model against all forecasting methods described in \autoref{forecasting_methods}. 
We select hyper-parameters for each baseline using time series cross-validation, which is similar to the test dataset. For choosing the optimal parameters, we use cross-validated RMSE.

\section{Methodology} \label{proposed_algorithm}

\subsection{Data preparation} \label{data_preparation}

\paragraph{Clustering}
Entities in the datasets may share similar trend patterns if they have a similar time series structure. Thus, the algorithm may recommend similar models within the same cluster. We perform hierarchical clustering with a correlation metric, and we use the generated cluster labels in the sequential embedding construction as an additional feature.

\paragraph{Time series normalization}
For the financial dataset, we create four moving averages for the traded volume, the adjusted close, and the open. The moving averages cover periods 5, 10, 15, and 30 days and represent weekly and monthly trends. We generate five statistical features. They are mean, standard deviation, skewness, kurtosis, and the autocorrelation of close prices. Similarly, we include five trading indicators, RSI, ATR, OBV, Hilbert Transform period, and phase.

\paragraph{Time series to graph conversion} \label{graph_conversion}
In order for the model to learn the structure of the given time series data, firstly, the time series needs to be converted into the graph.
We start by applying the single spectrum analysis (SSA), \cite{ssa}, method to the time series. It provides a representation of a given time series in the form of eigenvalues and eigenvectors of a matrix composed of a time series. With the SSA results, we can build a visibility graph to map a time series into a graph, \cite{visbi}. The benefits are that every node can see at least its closest neighbors, we do not define the direction in the links and the visibility is resistant to vertical and horizontal rescaling and translating.
We generate visibility graphs by designing visibility criteria. Two arbitrary time series values $(v_a, y_a)$ and $(v_b,y_b)$ will be visible to each other and therefore will be connected in the constructed graph, if any other data $(v_c,y_c)$ between them fulfils:
  
$$ y_c <y_b + (y_a-y_b)\frac{v_b-v_c}{v_b-v_a}$$.

The resulting visibility graphs will result in a $N \times N$ matrix of size $10^9$, as we build graphs for $M$ entities. Therefore, we make use of graph embeddings. Specifically, we use the random walk node2vec, \cite{RW}. 
The node2vec algorithm has two parameters, $Q$, which defines how probable the random walk would discover the undiscovered part of the graph, and $P$, which defines how probable the random walk would return to the previous node. The algorithm node2vec embedding using random walk works as follows:
  \begin{itemize}
    \item Compute random walk probabilities $P$ and $Q$
    \item Simulate $r$ random walks of length $l$ starting from each node $u$
    \item Optimize the node2vec objective using Stochastic Gradient Descent
  \end{itemize}
  
\subsection{Model architecture} \label{model_architecture}
\begin{figure}[ht]
  \centering
  \includegraphics[width=\columnwidth]{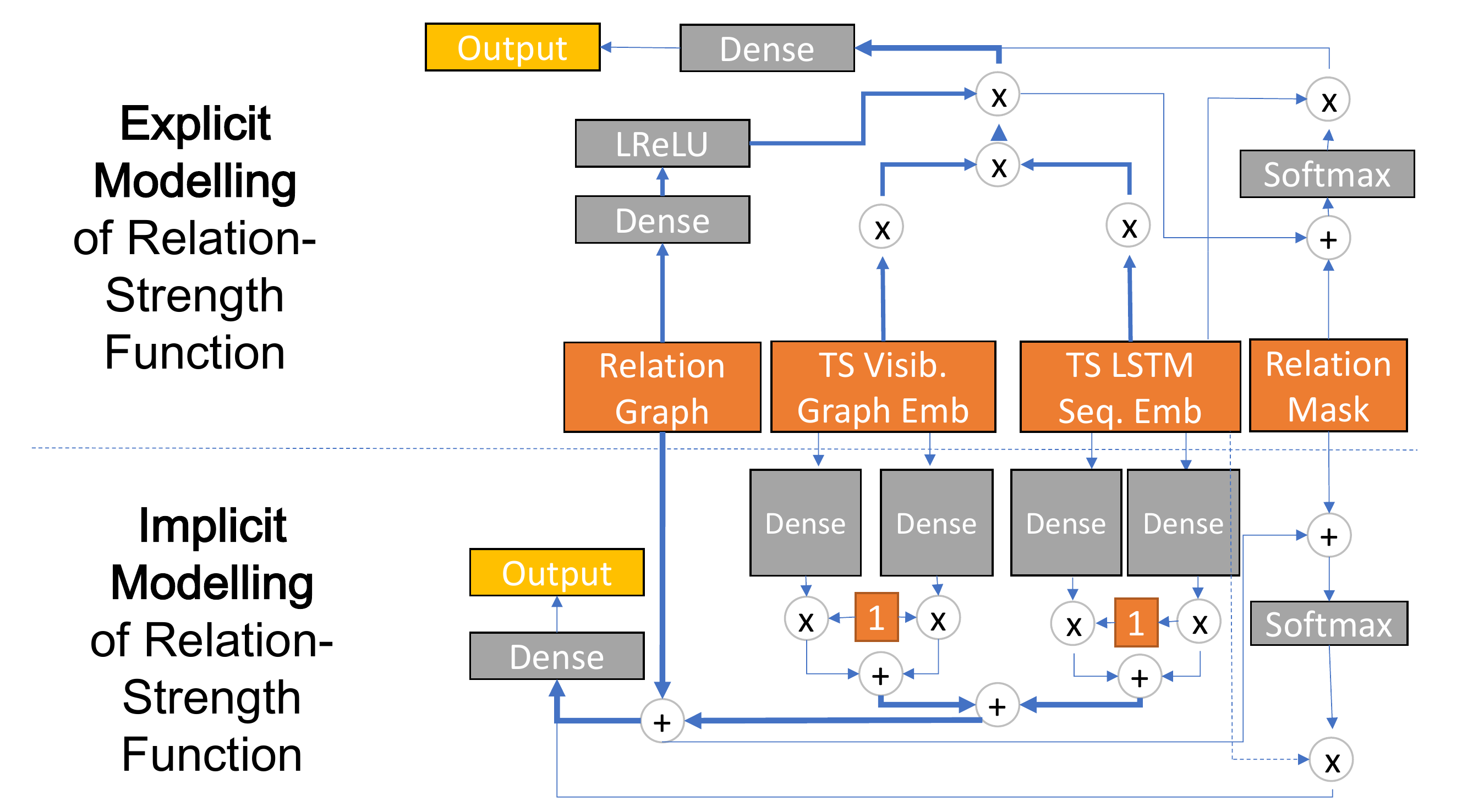}
  \caption{Algorithm Structure}
  \label{fig:validation_scheme}
\end{figure}

As illustrated in \autoref{fig:validation_scheme}, the framework contains three inputs, a relation graph, a time-series visibility graph embedding, and a time-series sequential embedding.

\paragraph{Relation graph}
We must consider how to model the influence between different entities, especially the ones with explicit relations. We can consider this as an injection of explicit domain knowledge, i.e., stock relations, into the data-driven approach for sequential embedding learning. For this, we use the multi-hot binary encoded matrix as an input.

\paragraph{Time series visibility graph embedding}
We transform each given Time Series into a visibility graph embedding.

\paragraph{Time series sequential embedding}
We must consider the strong temporal dynamics of the time series in our datasets. We apply a sequential embedding layer to capture the sequential dependencies in the historical data. Since RNN has achieved significant performance to process sequential data, \cite{RNN}, we opt for RNN to learn the sequential embeddings.
Among the different RNN, we choose LSTM due to its ability to capture long-term dependency.
We feed the historical time series data of entity $i$, as well as its constructed features, at time-step $t (X_t^t)$ to an LSTM network and take the last hidden state $(h_t^t)$ as the sequential embedding $(e_t^t)$ of stock, thus having
$$E^t = LSTM(X^t)$$, 
where $E^t \in R^{N \times U}$ denotes the sequential embeddings of all entities, and $U$ denotes the number of hidden units in LSTM.

Next, we describe two designs of the time-sensitive relation-strength function, which differ in whether to model the interaction between two entities explicitly or implicitly. We take the proposed architecture from \cite{TRR} and adapt it to take a time series graph embedding. 

\paragraph{Explicit Modeling} 
We define the relation strength function as:
\begin{align*}
    g(a_{ji}, e_i^t,e_j^t,e^{'t}_i,e^{'t}_j) = \\
    e^{t^T}_i e_j^t * e^{'t^T}_i e_j^{'t} * \phi (w^T a_{ji}+b),
\end{align*}
where $w$ and $b$ are the model parameters to be learned, $\phi$ is an activation function, $e_t$ is time series embedding, $e_{'t}$ is time series visibility graph embedding, $a_{ji}$-relation graph. Two terms determine the relation. The first term shows the similarity between the two entities at the moment of $t$. The hypothesis is that by being more similar at the current time, likely, their relations will impact the values. The second term is a nonlinear regression model on the relations, where each element in $w$ denotes the weights, and $b$ is a bias term. 

\paragraph{Implicit Modeling} 
We define the relationship strength by feeding the sequential and visibility graph embeddings and the relation vector into a fully connected layer as following:
\begin{align*}
g(a_{ji}, e_i^t,e_j^t,e^{'t}_i,e^{'t}_j) = \\
\phi (w^T[e_i^{t^T},e_j^{t^T},e_i^{'t^T},e_j^{'t^T},a_{ji}^T ]+b),
\end{align*}

where $w$ and $b$ are the model parameters to be learned, $\phi$ is an activation function, $e_t$ is time series embedding, $e_{'t}$ is the time series visibility graph embedding, and the $a_{ji}$-relation graph. The outputs are normalized using a softmax function.

\section{Hypotheses} \label{hypotheses}
We want to understand whether the proposed model is useful by evaluating memory usage, training, and processing time and error metric. For this, we set the following list of hypotheses:

\begin{itemize}
  \item The proposed algorithm performs better than LSTM without graph relation and visibility graph embedding.
  \item The proposed algorithm performs better than random model selection on all datasets.
  \item The proposed algorithm performs faster than Cross-Validation on all datasets while having the performance close to the CV on at least one data set.
  \item The proposed algorithm has accuracy on par with SOTA baselines in most of the cases.
  \item The proposed algorithm recommends the models more accurately than a meta-learning approach.
  \item The recommended models from the proposed algorithm give better prediction than the probability model selection approach. 
  \item It is possible to identify in which cases it is better to use implicit or explicit modeling.
\end{itemize}

\section{Experiments} \label{Experiments}
\begin{figure}[ht]
  \centering
  \includegraphics[width=\columnwidth]{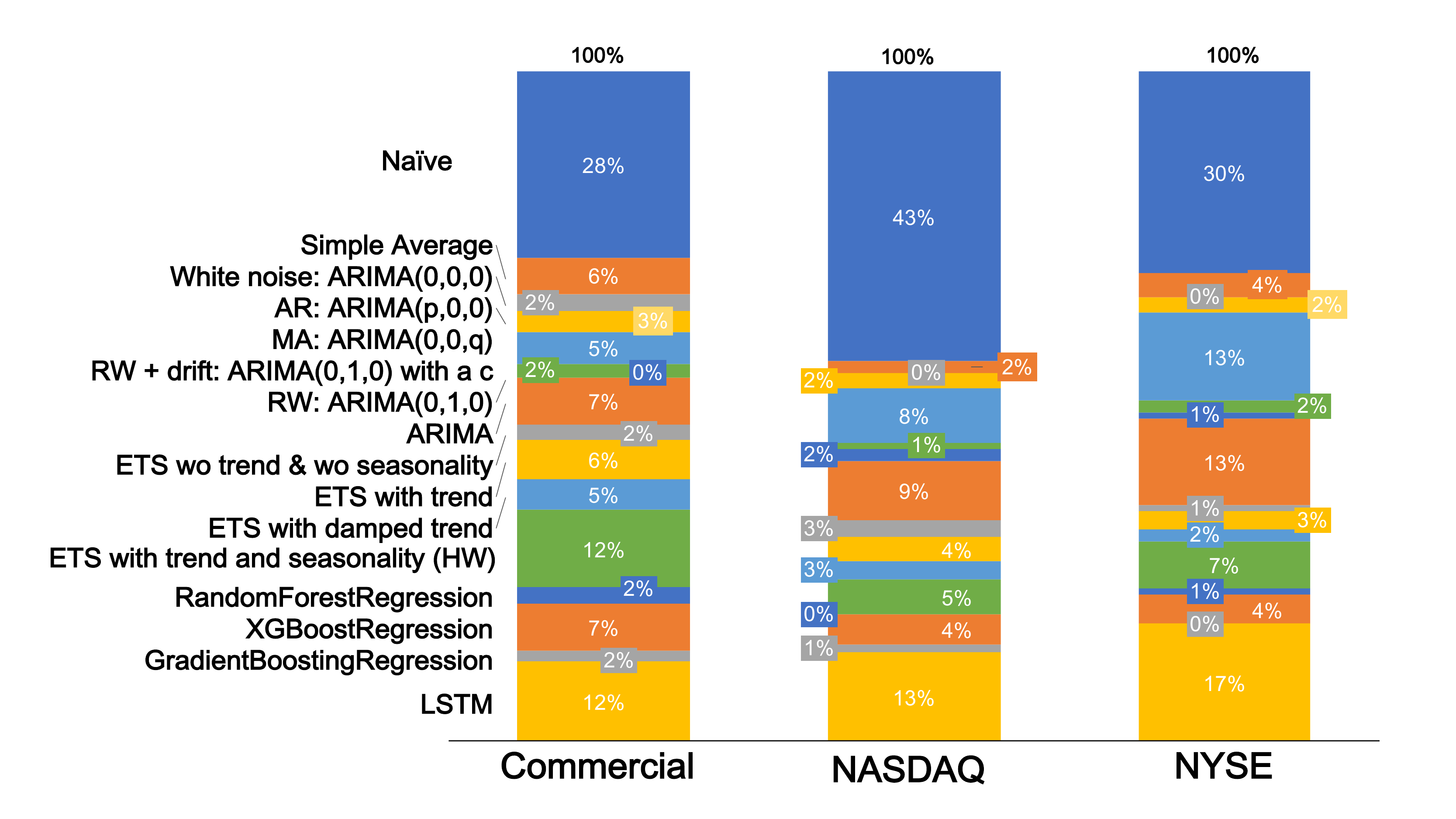}
  \caption{Best Model per Dataset}
  \label{fig:best_model}
\end{figure}

Before we compare our algorithm with the baselines, we need to label the time series with their "best" associated model. The labelled distribution is depicted in \autoref{fig:best_model}. The naive method is the most common "best" class, ranging from $28\%$ to $43\%$ for different datasets. Although part of the reason is that we perform only 1-day ahead forecasting. Therefore the next value might not change drastically. Similarly, the datasets are not highly volatile.

We compared against two baselines, model recommendation and order prediction. Random, CV, meta-learning, and penalized likelihood serve as the model recommendation baselines. Optimal hyperparameters were selected via CV on the same parameter grid, if applicable. 
Based on the results, we can conclude following the properties of the proposed model GNN. On all datasets, GNN with explicit relationships performed slightly better than GNN with the implicit relationship at the cost of increased computational time. Therefore the usage of explicit relationships is preferable. For the stock data, the GNN model has not shown noticeable improvements compared to a baseline approach. The LSTM embeddings do not seem to be ideal for the dataset. Nevertheless, SMAPE for stock data for GNN is on par with other methods.
Regarding the commercial dataset, GNN has performed similarly to the best model.
For the stock market, GNN has shown mediocre results in the model recommendation. The penalized likelihood approach greatly surpassed GNN. Meta-learning has also performed better. However, the SMAPE metric was worse. It shows that for some datasets, GNN cannot surpass the baselines. However, for commercial dataset, GNN has performed significantly better. 

\section {Discussion} \label{Discussion}

\subsection{Model recommendation classification accuracy comparison}
\begin{figure}
  \centering
  \includegraphics[width=\columnwidth]{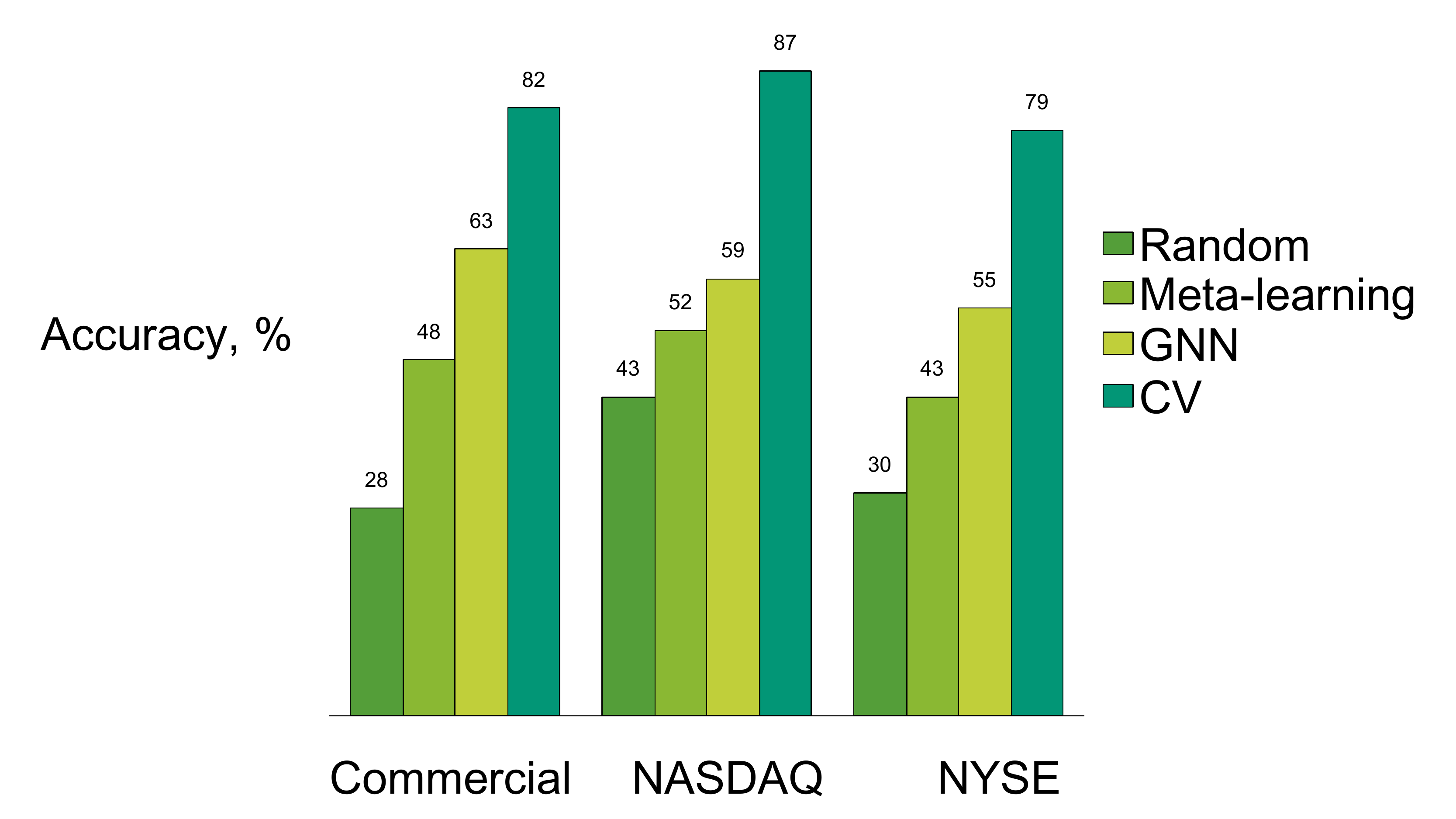}
  \caption{Classification Accuracy for the Baselines}
  \label{fig:classification_accuracy}
\end{figure}

In \autoref{fig:classification_accuracy}, we compare classification accuracy for different baselines. We define the "Random" baseline as the most common method. Namely, it will always pick the naive method. CV recommends the best model based on the exhaustive testing for all training periods. However, the best model on the testing period may sometimes be different from the models on the training different. Therefore this method does not reach the best accuracy.

The proposed algorithm performed better than meta-learning on all datasets. It was the primary baseline since the CV is quite hard to surpass. However, meta-learning is much faster to set up and run. Therefore both methods may be applicable in the industry setting.

\subsection{Dependency between sequential size and best models}
\begin{figure}[ht]
  \centering
  \includegraphics[width=\columnwidth]{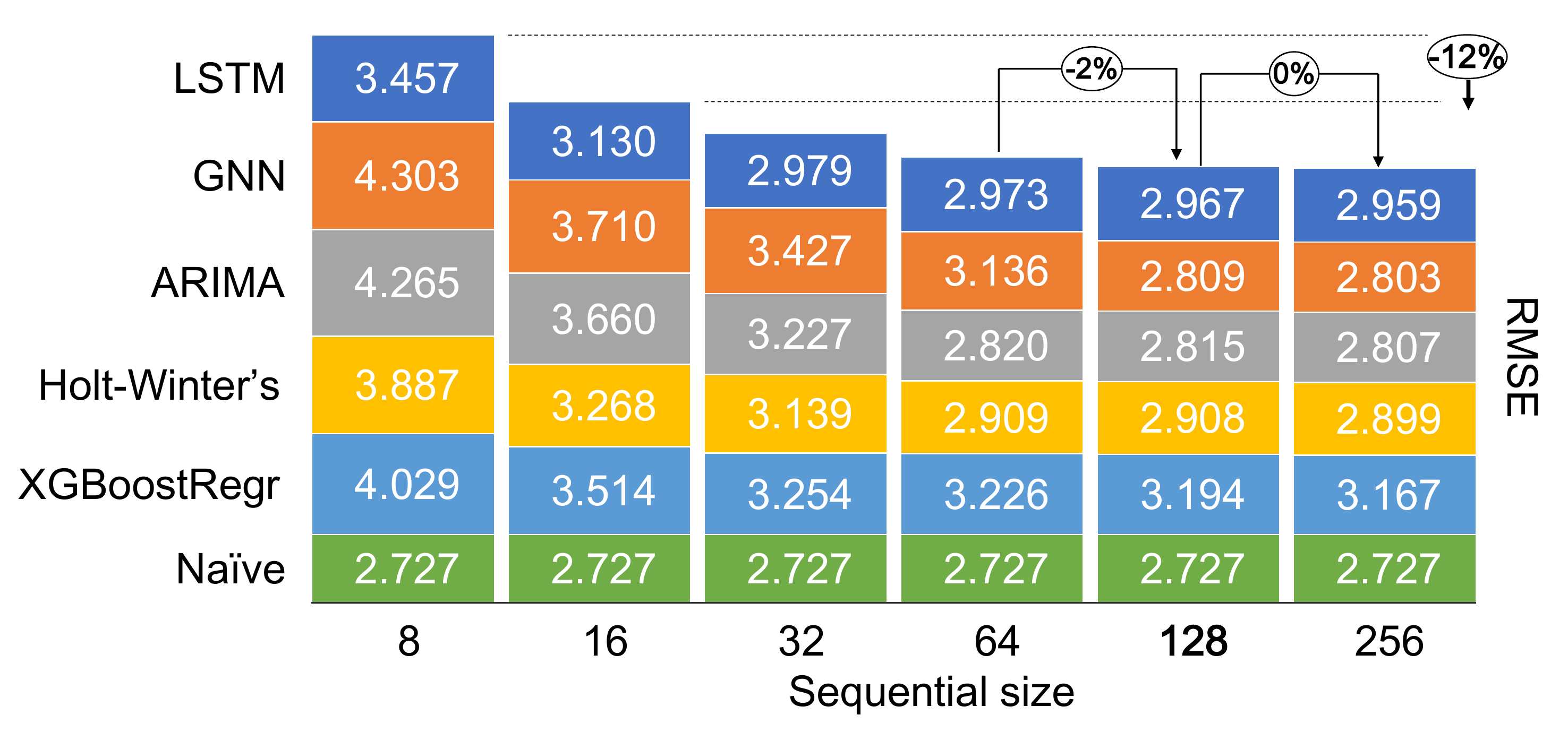}
  \caption{Dependency between Sequential Size and Best Models for the Commercial Dataset based on MSE}
  \label{fig:sequential_size}
\end{figure}

\autoref{fig:sequential_size} shows scores for the different training periods and corresponding models. We can notice a rapid improvement in scores (MSE) for most of the models until a sequential size of $128$. After that, both the improvements are imperceptible and becoming unmanageable to train. Therefore the chosen sequential size of the embeddings is equal to $128$. 

\subsection{Dependency between Accuracy and batch size }
\begin{figure}[ht]
  \centering
  \includegraphics[width=\columnwidth]{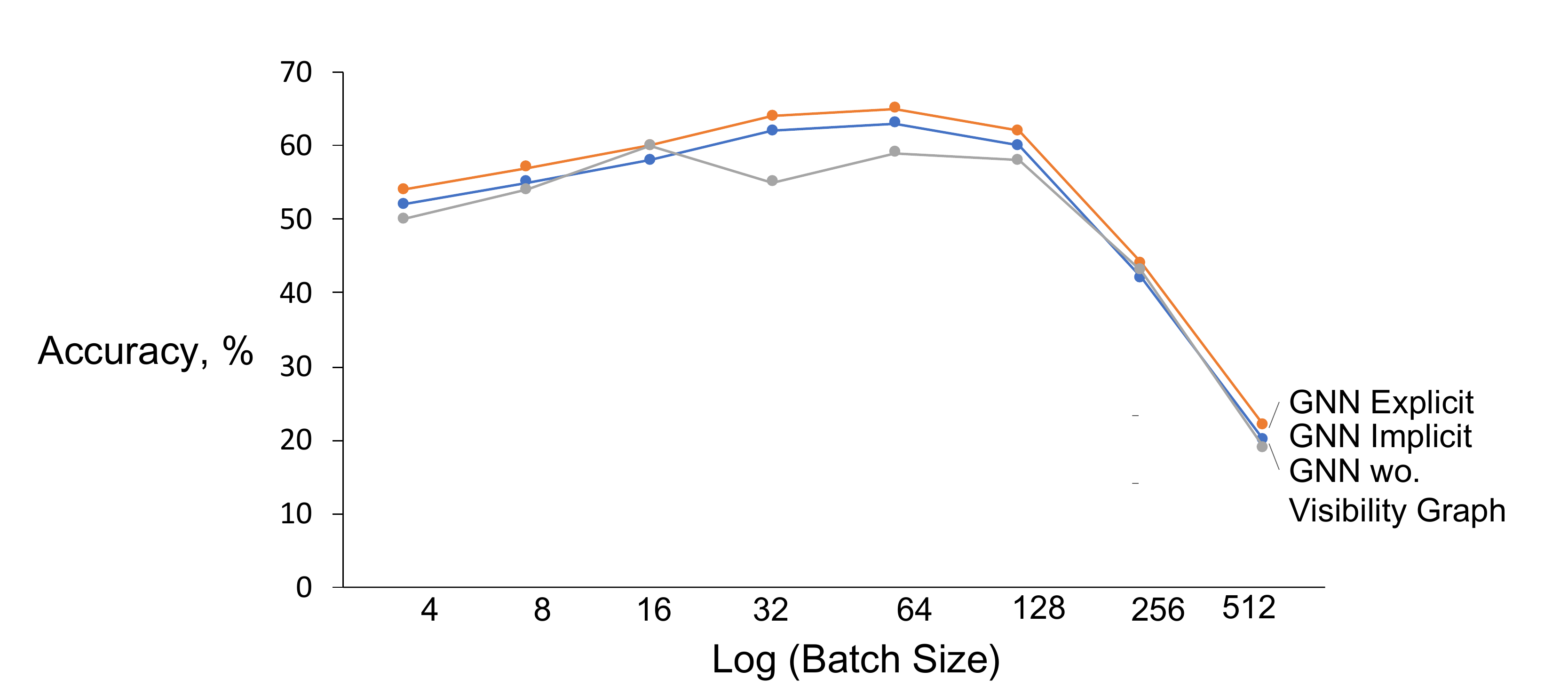}
  \caption{Dependency between Model Recommendation Accuracy and Log Batch Size for commercial dataset}
  \label{fig:log_accuracy}
\end{figure}

We can see the accuracy of all GNN models in \autoref{fig:log_accuracy}. One observation is that using a larger batch leads to a degradation in the quality of the model, as measured by its ability to generalize. Therefore for our study has a batch size of $64$. 

\subsection{Hyperparameter tuning}

We present the chosen hyperparameter setting on \autoref{Tab: hyperparameters}. Fifty epochs are usually enough for the loss to start converging. We set steps to 1. In the current study, we perform only 1-day ahead forecasting. We set the hidden unit size to 64. For the algorithm of time series to graph conversion, we set the following parameters after the experiments. SSA window equals $32$, $r$ is $16$, $l$ is $16$ and embedding size is $128$. 

\begin{table}
\centering
\caption{\textsc{Final Hyperparameters for the proposed GNN Model}\label{Tab: hyperparameters}}
\resizebox{0.48\textwidth}{!}{%
\begin{tabular}{lc}
\hline
Hyperparameter  & Value  \\ \hline
Relation Type                        & Explicit                      \\ \hline
Learning Rate                        & 0.01                        \\ \hline
SSA Window                          & 32                         \\ \hline
\# of Random Walks for Graph Embedding            & 16                         \\ \hline
Length of Random Walks for Graph Embedding          & 16                         \\ \hline
Graph Embedding Length                    & 128                        \\ \hline
Sequential Size                       & 128                        \\ \hline
Epochs                            & 50                         \\ \hline
Steps                            & 1                         \\ \hline
Batch Size                         & 32                         \\ \hline
Hidden Unit Size                       & 64                         \\ \hline
\end{tabular}
}
\end{table}

Based on the results of this study, we can establish the following guidelines. 
First, compare with other baselines to determine whether GNN is applicable for the current dataset and whether implicit or explicit relation is preferable.
Second, engineer features so that sequential embedding can bring more predictive power and have more data about structure about time series.
Third, construct a relation graph with a relation ratio of at least 5\%. The sparse relation graph will result in the loss of the predictive power of the LSTM embedding.
Fourth, perform extensive hyperparameter tuning since some untuned parameters can significantly limit predictive power.

\section{Summary} \label{summary}
We designed the present study to prove that the model using GNN can recommend appropriate models based on the time series structure. 
After quantitative experiments on commercial and public datasets, we can conclude that GNN proved its competitiveness on the commercial dataset. However, it has not shown significant results on the public dataset in terms of prediction power. 
Nevertheless, the model bested other baselines for both datasets. 

We presented the conversion of time series to a graph to construct the model, which can learn the underlying time series data structure. Further, we outlined two types of time-sensitive relation-strength functions to tie together three incoming inputs in the model. The model was configured to output both the next value prediction and model recommendation.

The model is suitable for small datasets since it shows adequate results, even with a small training period. However, it is not robust for the data with a small number of entities or relations between entities, since inside the architecture, it relies heavily on the generated relation graph.

\section{Limitations} \label{limitations}
We want to outline limitations to the presented approach.
First, the model surpasses the target baseline only on the commercial dataset. Therefore it is not the best choice for all possible datasets.
Second, the model requires careful and thorough hyperparameter tuning of several pipelines such as SSA, visibility graph, embedding, relation function, and NN architecture.
Third, the model has several complicated intermediary prepossessing steps that impede its usefulness in the industry setting, although we can mitigate this by constructing a ready-to-use solution. 
Fourth, the model is computationally expensive to use. Therefore it is not suitable for the "online" setting.
Fifth, as the model levers neural networks, it is "black-box" with limited interpretability.

\section{Future Work} \label{future_work}

There exist many ways for further improvement of the proposed algorithm. 
Firstly, we can improve the model by using corresponding features of each time series, such as product name embedding. The architecture has to be adapted to allow having the fourth additional input.

Second, we can use the "triple-barrier" method to transform the experiment into a model recommendation classification task. Thus, we can classify each time series to recommend the models, which are best suited for a particular time series. 
This way, it will be possible to improve the results on the public dataset. 

 Another way for future improvement is to extend forecasting from 1-day ahead to several day-ahead. Indeed, in the current study, the properties of many powerful algorithms were hidden since the naive method is quite powerful in 1-day ahead forecasting. However, in n-day ahead forecasting, other methods demonstrate predictive power, which will change the models' performance on the baselines.

\ifCLASSOPTIONcompsoc
 \section*{Acknowledgments}
\else
 \section*{Acknowledgment}
\fi

Sections I, II, III, IV were supported by the Ministry of Education  and  Science  of  the  Russian  Federation  (Grant no. 14.756.31.0001). Other sections were supported by the Mexican National Council for Science and Technology (CONACYT), 2018-000009-01EXTF-00154.


\bibliographystyle{IEEEtran}
\bibliography{bib}

\end{document}